\documentclass[12pt]{article}
\font\titlefont=cmbx10 scaled \magstep3
\begin{document}
\input{epsf}

\begin{center}
{\titlefont Spacetime in Semiclassical Gravity}

\vskip .3in
L.H. Ford\footnote{email: ford@cosmos.phy.tufts.edu} \\
\vskip .1in
Institute of Cosmology,
Department of Physics and Astronomy\\
Tufts University\\
Medford, Massachusetts 02155\\
\end{center}

\vskip .2in

\begin{abstract}
This article will summarize selected aspects of the semiclassical theory of
gravity, which involves a classical gravitational field coupled to quantum
matter fields. Among the issues which will be discussed are the role of
quantum effects in black hole physics and in cosmology, the effects of
quantum violations of the classical energy conditions, and inequalities 
which constrain the extent of such violations. We will also examine
the first steps beyond semiclassical gravity, when the effects of 
spacetime geometry fluctuations start to appear.
\end{abstract}

\baselineskip=14pt

\section{Introduction} 

  This article will deal with the semiclassical approximation, in which
the gravitational field is classical, but is coupled to quantum matter fields.
The semiclassical theory consists of two aspects: (1) Quantum field theory
in curved spacetime and (2) The semiclassical Einstein equation.
Quantum field theory in curved spacetime describes the effects of gravity 
upon the quantum fields. Here a number of nontrivial effects arise,
including particle creation, negative energy densities, and black hole
evaporation. The semiclassical Einstein equation describes how quantum
fields act as the source of gravity.  This equation is usually taken to
be the classical Einstein equation, with the source as the quantum 
expectation value of the matter field stress tensor operator, that is
\begin{equation}
G_{\mu\nu} = 8 \pi \langle T_{\mu\nu} \rangle \,.  \label{eq:sce}
\end{equation}
This expectation value is only defined after suitable regularization and
renormalization.

In this article, we will use units (Planck units) in which Newton's constant, 
the speed of light, and $\hbar$ are set to one: $G = c = \hbar = 1$.     
This makes all physical quantities dimensionless. Thus masses, lengths, and
times are expressed as dimensionless multiples of the Planck mass,
$m_P = \sqrt{\hbar c/G} = 2.2 \times 10^{-5} {\rm g}$, the Planck length,
$\ell_P =\sqrt{\hbar G/c^3} = 1.6 \times 10^{-33} {\rm cm}$, and the Planck
time, $t_P = \sqrt{\hbar G/c^5} = 5.4 \times 10^{-44} {\rm s}$, respectively.

\section{Renormalization of $\langle T_{\mu\nu} \rangle$}
\label{sec:renorm}

Here we will outline of the procedure for extracting a meaningful,
finite part from the formally divergent expectation value of the stress tensor.
More detailed accounts can be found in the books by Birrell and 
Davies\cite{BD} and by Fulling\cite{Fulling} .
The first step is to introduce a formal regularization scheme, which
renders the expectation value finite, but dependent upon an arbitrary
regulator parameter. One possible choice is to separate the spacetime points
at which the fields in $ T_{\mu\nu}$ are evaluated, and then to average over
the direction of separation. This leaves $\langle T_{\mu\nu} \rangle$
depending upon an invariant measure of the distance between the two points.
This is conventionally chosen to be one-half of the square of the geodesic
distance, denoted by $\sigma$.  

The asymptotic form for the regularized expression in the limit of small
$\sigma$ can be shown to be  
\begin{equation}
\langle T_{\mu\nu} \rangle \sim A\: \frac{g_{\mu\nu}}{\sigma^2} + 
B\: \frac{G_{\mu\nu}}{\sigma} + \bigl(C_1 H^{(1)}_{\mu\nu} +
C_2 H^{(2)}_{\mu\nu} \bigr)\: \ln\,\sigma. \label{eq:reg}
\end{equation}
Here $A$, $B$, $C_1$, and  $C_2$ are constants, $G_{\mu\nu}$ is the Einstein 
tensor, and the $H^{(1)}_{\mu\nu}$ and $H^{(2)}_{\mu\nu}$ tensors are 
covariantly conserved tensors which are quadratic in the Riemann tensor.
Specifically, they are the functional derivatives with respect to the metric
tensor of the square of the scalar curvature and of the Ricci tensor, 
respectively:
\begin{eqnarray}
H^{(1)}_{\mu\nu} &\equiv& \frac{1}{\sqrt{-g}} \frac{\delta}{\delta g^{\mu\nu}} 
                                 \bigl[\sqrt{-g} R^2 \bigr] \nonumber \\
&=& 2\nabla_\nu \nabla_\mu R -2g_{\mu\nu}\nabla_\rho \nabla^\rho R
 - \frac{1}{2} g_{\mu\nu} R^2 +2R R_{\mu\nu},  \label{eq:H1}
\end{eqnarray}
and
\begin{eqnarray}
H^{(2)}_{\mu\nu} &\equiv& 
                \frac{1}{\sqrt{-g}} \frac{\delta}{\delta g^{\mu\nu}} 
            \bigl[\sqrt{-g} R_{\alpha\beta}R^{\alpha\beta} \bigr] 
= 2\nabla_\alpha \nabla_\nu R_\mu^\alpha - \nabla_\rho \nabla^\rho R_{\mu\nu}
\nonumber \\ &{}& -\frac{1}{2} g_{\mu\nu}\nabla_\rho \nabla^\rho R
  -\frac{1}{2}g_{\mu\nu} R_{\alpha\beta}R^{\alpha\beta} 
   +2R_\mu^\rho R_{\rho\nu}.    \label{eq:H2}
\end{eqnarray}

The divergent parts of $\langle T_{\mu\nu} \rangle$ may be absorbed by
renormalization of counterterms in the gravitational action. Write this
action as
\begin{equation}
S_G = \frac{1}{16\pi G_0} \int d^4x\,\sqrt{-g}\, \Bigl( R -2\Lambda_0
      + \alpha_0 R^2 + \beta_0 R_{\alpha\beta}R^{\alpha\beta} \Bigr).
\end{equation}
We now include a matter action, $S_M$, and vary the total action, 
$S= S_G + S_M$, with respect to the metric. If we replace the classical
stress tensor in the resulting equation by the quantum expectation value,
$\langle T_{\mu\nu} \rangle$, we obtain the semiclassical Einstein equation
including the quadratic counterterms:
\begin{equation}
G_{\mu\nu} +\Lambda_0 g_{\mu\nu} +\alpha_0 H^{(1)}_{\mu\nu}
 +\beta_0 H^{(2)}_{\mu\nu} = 8\pi G_0 \langle T_{\mu\nu} \rangle.
\end{equation}
We may remove the divergent parts of $\langle T_{\mu\nu} \rangle$ in 
redefinitions of the coupling constants $G_0$, $\Lambda_0$, $\alpha_0$, 
and $\beta_0$. The renormalized values of these constants are then the physical
parameters in the gravitational theory. After renormalization, $G_0$ is 
replaced by $G$, the renormalized Newton's constant, which is the value 
actually
measured by the Cavendish experiment. Similarly, $\Lambda_0$ becomes the
renormalized cosmological constant $\Lambda$, which must be determined by 
observation. This is analogous to any other renormalization in field
theory, such as the renormalization of the mass and charge of the electron
in quantum electrodynamics.

In any case, the renormalized value of $\langle T_{\mu\nu} \rangle$ is 
obtained by subtracting the  terms which are divergent in the coincidence
limit. However, we are free to perform additional finite renormalizations
of the same form. Thus, $\langle T_{\mu\nu} \rangle_{ren}$ is defined only
up to the addition of multiples of the four covariantly conserved, geometrical
tensors $g_{\mu\nu}$, $G_{\mu\nu}$, $H^{(1)}_{\mu\nu}$, and $H^{(2)}_{\mu\nu}$.
Apart from this ambiguity, Wald\cite{Wald2} has shown under 
very general assumptions
that $\langle T_{\mu\nu} \rangle_{ren}$ is unique. Hence, at the end of the
calculation, the answer is independent of the details of the regularization 
and renormalization procedures employed.

\section{The Stability Problem in the Semiclassical Theory}

The classical Einstein equation is a second order, nonlinear, differential
equation for the spacetime metric tensor, because the Einstein tensor involves
up to second derivatives of the metric. As a second order system of
hyperbolic equations, it possesses
a well-posed initial value formulation: if one specifies the metric and its
first derivatives on a spacelike hypersurface, there exists a unique solution
of the equations\cite{Wald}. 
This is the usual situation in physics, where the fundamental
equations can be cast as a second order system. (For example, Maxwell's
equations are equivalent to a set of second order wave equations for the
vector and scalar potentials.) 

There is a problem with the semiclassical Einstein equation in that it
is potentially a fourth-order system of equations. This arises from terms
involving second derivatives of the curvature tensor, and hence fourth
derivatives of the metric.  This leads to the unpleasant feature that a
unique solution would require specification of the metric and its first
three derivatives on a spacelike hypersurface. Even worse, it can lead to 
instability. The situation is analogous to that in classical electrodynamics
when radiation reaction in included in the equation of motion of a charged 
particle\cite{Jackson}. 
The Abraham-Lorentz equation, which includes the radiation
reaction force for a nonrelativistic particle, is third-order in time
and possesses runaway solutions. In electrodynamics, the problem is partially
solved by replacing the third-order Abraham-Lorentz equation by an
integrodifferential equation which is free of runaway solutions, but
exhibits acausal behavior on short time scales.  However, this acausality
is on a time scale small compared to the Compton time of the particle.
As such, it lies outside of the domain of validity of classical 
electrodynamics. 

Several authors\cite{HW,PS,AMM03}
 have discussed the instability problem in semiclassical
gravity theory. Some  of the proposed resolutions of this problem involve
reformulating the theory to eliminate unstable solutions (analogous
to the integrodifferential equation in electrodynamics), or regarding
the semiclassical theory as valid only for spacetimes which pass a stability
criterion. These are sensible approaches to the issue. Basically, one wishes 
to have a theory which can approximately describe the backreaction of
quantum fields on scales well above the Planck scale. It is important 
to keep in mind that the semiclassical theory is an approximation which
must ultimately fail in situations where the quantum nature of gravity
itself plays a crucial role.

\section{The Hawking Effect}

One of the great successes of quantum field theory in curved spacetime
and of semiclassical gravity is the elegant connection between black
hole physics and thermodynamics forged by the Hawking effect. Classical
black hole physics suffers from Bekenstein's paradox\cite{Bekenstein}: 
one could
throw hot objects into a black hole and apparently decrease the net entropy 
of the universe. This paradox can be resolved by assigning an entropy
to a black hole which is proportional to the area of the event horizon.
Hawking\cite{Hawking} 
carried this reasoning one step further by showing that black
holes are hot objects in a literal sense and emit thermal radiation.
The outgoing radiation consists of particles quantum mechanically created
in a region outside of the event horizon, and carries away energy and
entropy from the black hole. The resulting decrease in mass of the hole
arises from a steady flux of {\it negative energy} into the horizon,
and is consistently described by the semiclassical Einstein equation,
Eq.~(\ref{eq:sce}), so long as the black hole's mass is well above
the Planck mass.

Although the Hawking effect provides an elegant unification of
thermodynamics, gravity and quantum field theory, there are still
unanswered questions. One is the ``information puzzle'', the issue
of whether information which goes into the black hole during its
semiclassical phase can be recovered. Hawking\cite{Hawking76} originally
proposed that this information is irrevocably lost and that black
hole evaporation is not described by a unitary evolution. This view
has been disputed by several other physicists\cite{tHooft}, 
who have argued that
a complete quantum mechanical description of the evaporation process
should be unitary. More recently, Hawking\cite{Hawking04} 
has agreed with this view.
However, even if the evolution is unitary, the details by which information
is recovered are still unclear. One possibility is that the outgoing radiation
is not exactly thermal, but contains some subtle correlations which
carry the information about the details of the matter which fell into
the black hole. If this suggestion is correct, it is not clear
just how these correlations arise.

A second mystery raised by the Hawking effect is the ``tranplanckian
problem''.  This problem arises because the modes which will eventually
become populated with the outgoing thermal radiation start out with
extremely high frequencies before the black hole formed. These modes
enter the collapsing body and then exit just before the horizon forms,
undergoing an enormous redshift. However, as they enter and pass through the
body, their frequencies are vastly higher than the Planck scale. If one
postulates that full quantum gravity will impose an effective cutoff at 
the Planck scale, then there seems to be a conflict; a cutoff at any
reasonable frequency would eliminate the modes needed for the Hawking
radiation. For a black hole of mass $M$ to evaporate, one needs to start
with modes whose frequency is of order
\begin{equation}
\omega \approx \frac{{\rm e}^{M^2}}{M} \,.
\end{equation}
For a stellar mass size black hole, this corresponds to 
$\omega \approx 10^{10^{75}}\, {\rm g}$, which is vastly larger than the 
mass of the observable universe.
One possible resolution\cite{Jacobson,Unruh95,CJ96} 
of this problem is to postulate a modified
dispersion relation which allows for ``mode creation'', whereby
the modes would appear shortly before they are needed to carry the
thermal radiation. However, this solution will require new microphysics,
including breaking of local Lorentz invariance.

\section{Quantum Effects in the Early Universe}

It is likely that there is a period in the history of the universe
during which quantum effects are important, but one is sufficiently
far from the Planck regime that a full theory of quantum gravity is
not needed. In this case, the semiclassical theory is applicable.
Among the quantum effects expected in an expanding universe is
quantum particle creation\cite{Parker}.
Inflationary models with inflation occurring
at scales below the Planck scale are plausible models for the early universe
in which semiclassical gravity should hold. Indeed, such models predict that
the density perturbations which later grew into galaxies had their
origins as quantum fluctuations during the inflationary 
epoch\cite{MC81,GP82,Hawking82,Starobinsky82,BST83}. This leads to the
remarkable prediction that the large scale structure of the present day
universe had its origin in quantum fluctuations of a scalar inflaton 
field. More precisely, quantum fluctuations of a nearly massless scalar
field in deSitter spacetime translate into an approximately scale invariant
spectrum of density perturbations. This picture seems to be consistent
with recent observations of the cosmic microwave background 
radiation\cite{MG04}.

\section{The Dark Energy Problem}

There is now strong evidence that the expansion of the present day
universe is accelerating. This evidence came first from observations
of type Ia supernovae\cite{Schmidt,Perlmutter}. This acceleration could 
be due to a nonzero value for the cosmological constant, but other 
possibilities are consistent with the observational data. These possibilities
go under the general term ``dark energy'', and require a negative pressure
whose magnitude is approximately equal to the energy density. It has
sometimes been suggested that the dark energy could be viewed as due
to quantum zero point energy.
 However, there are
some serious difficulties with this viewpoint. If we adopt the convention
renormalization approach discussed in Sect.~\ref{sec:renorm}, then the
renormalized value of the cosmological constant $\Lambda$ is completely
arbitrary. At this level, quantum field theory in curved spacetime can
no more calculate $\Lambda$ then quantum electrodynamics can calculate
the mass of the electron. We could take a more radical approach and seek
some physical principle which effectively fixes the value of the regulator
parameter to a definite, nonzero value. However, for the first term on
the right hand side of  Eq.~(\ref{eq:reg}) to be the dark energy, we would
have to take $\sigma \approx (0.01 {\rm cm})^2$. It is very hard to imagine 
what new physics would introduce a cutoff on a scale of the order of
$0.01 {\rm cm}$.

There is still a possibility that the dark energy could be due to 
some more complicated mechanism which involves quantum effects. One
appealing idea is that there might be a mechanism for the cosmological 
constant to decay from a large value in the early universe to a smaller, but
nonzero value today. Numerous authors
\cite{Dolgov,F85,F87,Weinberg,DEZ,TW,AW01,Schutzhold,F02} have discussed 
models for the decay of the cosmological constant, or models which otherwise
attribute a quantum origin to the dark energy\cite{SH98,PR99}. However, at the
present time there is no widely accepted model which successfully links
dark energy with quantum processes.

\section{Negative Energy Density  for Quantum Fields}

One crucial feature of quantum matter fields as a source of gravity
is that they do not always satisfy conditions obeyed by known forms
of classical matter, such as positivity of the local energy density.
 Negative energy densities and fluxes arise even in flat spacetime.
A simple example is the Casimir effect\cite{Casimir}, 
where the vacuum state of the 
quantized electromagnetic field between a pair of perfectly 
conducting plates separated
by a distance $L$ is a state of constant negative energy density
\begin{equation}
\rho = \langle T_{tt} \rangle = - \frac{\pi^2}{720 L^4}.
\end{equation}
Even if the plates are not perfectly conducting, it is still possible to 
arrange for the energy density at the center to be negative\cite{SF02}.

Negative energy density can also arise as the result of quantum coherence
effects.  In fact, it may be shown under
rather general assumptions that quantum field theories admit states
for which the energy density will be negative somewhere\cite{EGJ,Kuo97}.
In simple cases, such as a free scalar field
in Minkowski spacetime, one can find states in which the energy density
can become arbitrarily negative at a given point.

We can illustrate the basic phenomenon of negative energy arising from
quantum coherence with a very simple example. Let 
the quantum state of the system be a superposition of the vacuum and a 
two particle state:
\begin{equation}
|\Psi\rangle = 
\frac{1}{\sqrt{1+\epsilon^2}} (|0\rangle + \epsilon|2\rangle). 
                                              \label{eq:vacplus2}
\end{equation}
Here we take the relative amplitude $\epsilon$ to be a real number. Let
the energy density operator be normal-ordered:
\begin{equation}
\rho = :T_{tt}:\, ,
\end{equation}
so that $\langle0|\rho|0\rangle =0.$ Then the expectation value of the energy
density in the above state is
\begin{equation}
\langle \rho \rangle = \frac{1}{1+\epsilon^2} 
      \Bigl[2 \epsilon {\rm Re}(\langle0|\rho|2\rangle)
            + \epsilon^2 \langle2|\rho|2\rangle \Bigr]\, .
\end{equation}
We may always choose $\epsilon$ to be sufficiently small that the first
term on the right hand side dominates the second term. However, the former
term may be either positive or negative. At any given point, we could choose
the sign of $\epsilon$ so as to make $\langle \rho \rangle  < 0$ at that point.
This example is a limiting case of a more general class of quantum states
which may exhibit negative energy densities, the squeezed states.

Note that the integral of $\rho$ over all space is the Hamiltonian, which 
does have non-negative expectation values:
\begin{equation}
\langle H \rangle = \int d^3 x \langle \rho \rangle \geq 0.
\end{equation}
In the above {\it vacuum + two particle} example, the matrix element 
$\langle0|\rho|2\rangle$, which gives rise to the negative energy density,
has an integral over all space which vanishes, so only $\langle2|\rho|2\rangle$
contributes to the Hamiltonian.

\section{Some Possible Consequences of Quantum Violation of Classical
Energy Conditions}

The existence of negative energy density can give rise to a number of
 effects in which the predictions of semiclassical gravity
differ significantly from those of classical gravity theory. 

\subsection{Singularity Avoidance}

In the 1960's, several elegant theorems were proven by Penrose, Hawking,
and others\cite{HE} 
which demonstrate the inevitability of singularity formation
in gravitational collapse described by classical relativity. These singularity
theorems imply that the curvature singularities found in the exact solutions
for black holes or  for cosmological models are generic and signal a breakdown
of classical relativity theory. However, this does not tell us whether a full
quantum theory of gravity is needed to give a physically consistent, that is,
singularity free, picture of the end state of gravitational collapse or the 
origin of the universe.

A crucial feature of the proofs of the singularity theorems is the assumption
of a classical energy condition. There are several such conditions that
can be used, but a typical example is the weak energy condition. This states
that the stress tensor $T_{\mu\nu}$ must satisfy 
$T_{\mu\nu}\, u^\mu\,u^\nu \geq 0$ for all timelike vectors $u^\mu$. Thus
all observers must see the local energy density being non-negative.
It is not hard to understand why there could not be a singularity theorem
without an energy condition: the Einstein tensor $G_{\mu\nu}$ is a function
of the metric and its first two derivatives. Thus, every twice-differentiable
metric is a solution of the Einstein equation, $G_{\mu\nu} = 8 \pi T_{\mu\nu}$
for some choice of $T_{\mu\nu}$. We can also understand the role which
the weak energy and related conditions play. Positive energy density will
generate an attractive gravitational field and cause light rays to focus.
Once gravitational collapse has proceeded beyond a certain point, the
formation of a singularity is inevitable as long as gravity remains attractive.
The way to circumvent this conclusion is with exotic matter, such as
negative energy density, which can cause repulsive gravitational effects.

Given that quantum fields can violate the classical energy conditions,
there is a possibility that the semiclassical theory can produce realistic,
nonsingular black hole and cosmological solutions. This is a topic which has 
been investigated by several authors\cite{PF73,Saa,HK01}. However, it is
difficult to avoid having the curvature reach Planck dimensions before
saturating. In this case, the applicability of the semiclassical theory
is questionable. It is possible to avoid this difficulty with a carefully
selected quantum states\cite{PF73}, a nonminimal scalar field which violates
the energy conditions at the classical level\cite{Saa}, or by going to
models where gravity itself is quantized\cite{HK01}.

\subsection{Creation of Naked Singularities}

There is an opposite effect which might be caused by negative energy:
the {\it creation} of a naked singularity. The singularities formed in
gravitational collapse in classical relativity tend to be hidden from the
outside universe by event horizons. Penrose\cite{Penrose} 
has made a ``cosmic censorship
conjecture'' to the effect that this must always be the case. This implies 
that the breakdown of predictability caused by the singularity is limited
to the region inside the horizon. It is not yet 
known whether this conjecture is true, even in the context of classical
relativity with classical matter, obeying classical energy conditions. 
However, unrestricted negative energy would allow a counterexample to
this conjecture. The Reissner-Nordstr{\"o}m solution of Einstein's equation
describes a black hole of mass $M$ and electric charge $Q$. However, these 
black hole solutions have an upper limit on the electric charge in relation 
to the mass of $Q \leq M$ (in our units). There are solutions for 
which $Q > M$, but these describe a naked singularity. Simple classical
mechanisms for trying to convert a charged black hole into a naked
singularity fail. If we try to increase the charge of a black hole, the 
work needed to overcome the electrostatic repulsion causes the black
hole's mass to increase at least as much as the charge and keep $Q \leq M$.
However, unrestricted negative energy would offer a way to violate
cosmic censorship and create a naked singularity. We could shine a beam of
negative energy involving an uncharged quantum field into the black hole,
decrease $M$ without changing $Q$, and thereby cause a naked singularity
to appear\cite{FR90,FR92}.

\subsection{Violation of the Second Law of Thermodynamics?}

If it is possible to create unrestricted beams of negative energy, then
the second law would seem to be in jeopardy. One could shine the beam
of negative energy on a hot object and decrease its entropy without a
compensating entropy increase elsewhere. The purest form of this experiment
would involve shining the negative energy on a black hole. If the negative
energy is carried  by photons with wavelengths short compared to the size
of the black hole, it will be completely absorbed. That is, there will be
no backscattered radiation which might carry away entropy. Then the black
hole's mass, and hence its entropy, will decrease in violation of the
second law\cite{F78}.

\subsection{Traversable Wormholes and Warp Drive Spacetimes}

As noted above, virtually any conceivable spacetime is a solution of Einstein's
equation with some choice for the source. If the source violates the classical
energy conditions, some bizarre possibilities arise. An example are the 
traversable wormholes of Morris, Thorne and Yurtsever\cite{MTU}. 
These would function as tunnels
which could connect otherwise widely separated regions of the universe
by a short pathway. An essential requirement for a wormhole is exotic matter
which violates the weak energy condition. The reason for this is that light
rays must first enter one mouth of the wormhole, begin to converge and later
diverge so as to exit the other mouth of the wormhole without coming to a
focal point. In other words,the spacetime inside the wormhole must act
like a diverging lens, which can only be achieved by exotic matter.

The existence of traversable wormholes would be strange enough, but they have
an even more disturbing feature: they can be manipulated to create a
time machine\cite{Visser}. 
If the mouths of a wormhole move relative to one another, it
is possible for the resulting spacetime to possess ``closed timelike paths''.
On such a path, an observer could return to the same point in space and in
time, and by speeding up slightly, arrive at the starting point before leaving.
Needless to say, this would turn physics as we currently understand it on
its head and open the door to disturbing causal paradoxes.

An equally bizarre possibility was raised by Alcubierre\cite{Alcubierre},
who constructed a spacetime that functions as science fiction style
``warp drive''. It consist of a bubble of flat spacetime surrounded by
expanding and contracting regions imbedded in an asymptotically flat 
spacetime. The effect of the expansion and contraction is to cause the
bubble to move faster than the speed of light, as measured by a distant 
observer, even though locally everything moves inside the lightcone.
Again, negative energy is essential for the existence of this spacetime.    
  
\section{Quantum Inequalities}

It is clear that unrestrained violation of the classical energy conditions
would create major problems for physics. However, it is also clear that
quantum field theory does allow for some violations of these conditions.
This leads us to ask if there are constraints on negative energy density
in quantum field theory. The answer is yes; there are inequalities which
restrict the magnitude and duration of the negative energy seen by any
observer, known as {\it quantum inequalities}
\cite{F78,F91,FR95,FR97,FLAN,PF971,PFGQI,FE,Fewster}. 
In four spacetime dimensions,
a typical inequality for a massless field takes the form\cite{FR95,FR97,FE}
\begin{equation}
\int \rho(t)\, g(t)\, dt \geq -\frac{c}{t_0^4} \,.     \label{eq:QI}
\end{equation}
Here $\rho(t)$ is the energy density measured in the frame of an inertial
observer, $g(t)$ is a sampling function with characteristic width $t_0$,
and $c$ is a numerical constant which is typically less than one. The value
of $c$ depends upon the form of  $g(t)$ (e.g. Gaussian versus Lorentzian).  
The sampling function
and its width can be chosen arbitrarily, subject to some differentiability
conditions on $g(t)$. The essential message of an inequality such 
Eq.~(\ref{eq:QI}) is that there is an inverse relation between the duration
and magnitude of negative energy density. In particular, if an observer
sees a pulse of negative energy density with a magnitude of order $\rho_m$
lasting a time of order $\tau$, then we must have $\rho_m < 1/\tau^4$. 

Furthermore, that negative energy cannot arise in isolation, but must be 
accompanied by compensating positive energy. This fact, plus the quantum
inequalities, place very severe restrictions on the physical effects
which negative energy can create. Here is a brief summary of the implications
of quantum inequalities for some of the possible effects listed above.

\subsection{Violations of the Second Law and of Cosmic Censorship}

If we were to shine a pulse of negative energy onto a black hole so
as to decrease its entropy and violate the second law, the entropy decrease
would have to last long enough to be macroscopically observable. At a
minimum, it should be sustained for a time longer than the size of the
event horizon. If the negative energy is constrained by an inequality
of the form of Eq.~(\ref{eq:QI}), then it can be 
shown\cite{F78} that the resulting
entropy decrease is of the order of Boltzmann's constant or less. This
represents an entropy change associated with about one bit of information,
hardly a macroscopic violation of the second law.

The attempt to create a naked singularity by shining a pulse of negative
energy on an extreme, $Q = M$, charged black hole is similarly constrained.
Again, any naked singularity which is formed should last for a time long
compared to $M$. However, it can be argued\cite{FR90,FR92} 
that the resulting change in
the spacetime geometry may be smaller than the natural quantum fluctuations
on this time scale. Thus it seems that negative energy which obeys the
quantum inequality restrictions cannot produce a clear, unambiguous
violation of cosmic censorship.

\subsection{Constraints on Traversable Wormholes and Warp Drive}

The simplest quantum inequalities, such as Eq.~(\ref{eq:QI}) have been proven 
only in flat spacetime, and hence do not immediately apply to curved
spacetime. There is, however, a limiting case in which they can also be
used in curved spacetime. This is when the sampling time $t_0$ , as measured
in a local inertial frame, is small compared to the local radii of curvature
of the spacetime in the same frame. This means that the spacetime is 
effectively flat on the time scale of the sampling, and the flat space
inequality should also apply to curved spacetime. In the special cases
where explicit curved spacetime inequalities have been derived, they are 
consistent with this limit. That is, they reduce to the corresponding flat 
space inequality in the short sampling time limit. 

Even in the small $t_0$ limit, it is possible to put very strong restrictions
on the geometry of traversable wormholes and warp drives\cite{FR96}. 
The constraints
on wormhole geometries vary from one model to another. In some cases, the
throat of the wormhole is limited to be close to Planck dimensions, 
presumably outside of the domain of validity of semiclassical gravity. 
In other cases, the restrictions are slightly less severe, but still
require some length scales to be much smaller than others, such as a band of
negative energy no more than $10^{-13} {\rm cm}$ thick to support a wormhole 
with a $1 {\rm m}$ throat. This does not quite rule out all possible
wormholes based upon semiclassical gravity, but makes it hard to imagine
actually constructing one. Similar, very strong restrictions are placed
on warp drive spacetimes\cite{PF97b,ER97}, such as the Alcubierre model.  
     
\section{Beyond Semiclassical Gravity: Fluctuations}

The first extension of semiclassical gravity arises when we consider
fluctuations of the gravitational field. These can be due to two
causes: the quantum nature of gravity itself (active fluctuations)
and quantum fluctuations of the stress tensor (passive fluctuations).
The extension of the semiclassical theory to include fluctuations
is sometimes called stochastic gravity\cite{Moffat,MV99,HV03,HV04}. 
One of the criteria for the validity of the semiclassical theory based upon
Eq.~(\ref{eq:sce}) must be that fluctuations are small\cite{KF93}. 
This theory can break 
down even well above the Planck scale if the stress tensor fluctuations
are sufficiently large. A simple example is a quantum state which is a
superposition of two states, each of which describe a distinct classical
matter distribution (e.g. a $1000 {\rm kg}$ mass on one or the other 
side of a room).
Equation~(\ref{eq:sce}) predicts a gravitational field which is an average
of the fields due to the two distributions separately, 
(the effect of two $500 {\rm kg}$ masses
on opposite sides of the room). However, an actual measurement of the 
gravitational field should yield that of a single $1000 {\rm kg}$ mass,
but in different locations in different trials.

A treatment of small fluctuations of the gravitational field offers a
window into possible extensions beyond strict semiclassical gravity.
First we should be clear about the operational meaning of fluctuations
of gravity. A classical gravitational field or spacetime geometry
can be viewed as encoding all possible motions of test particles in that
geometry. Consequently, fluctuations of spacetime imply Brownian motion
of the test particles, which can be characterized by mean squared
deviations from classical geodesics. 

\begin{figure}
\begin{center}
\leavevmode\epsfysize=8cm\epsffile{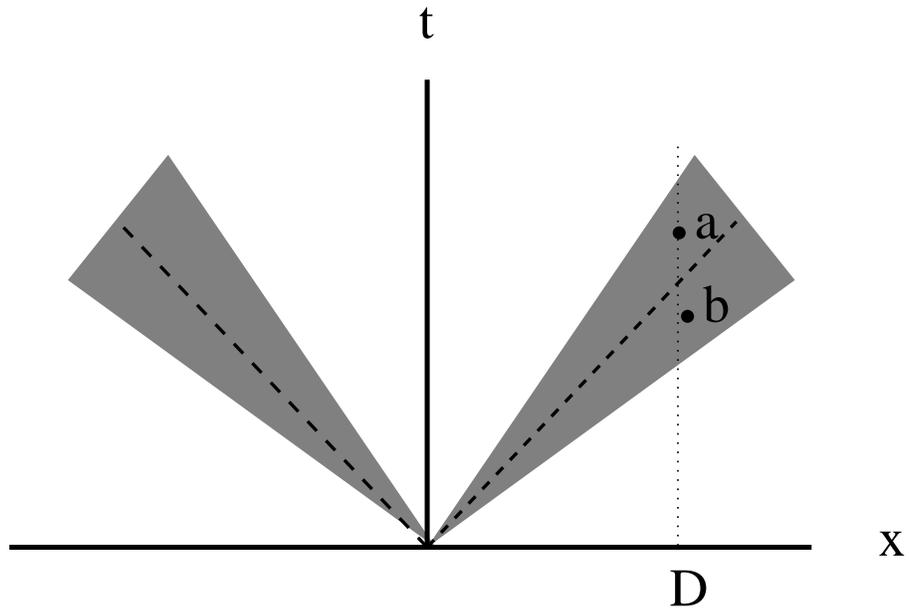}
\end{center}
\caption{The effects of lightcone fluctuations are illustrated. The dashed
lines represent the average lightcone. However, pulses which are emitted
at the origin can arrive at the worldline of a detector (vertical dotted
line) a mean distance $D$ away at different times. A pulse detected at
point $a$ travels slower than the mean speed of light, but one detected
at point $b$ has traveled fasted than the mean speed of light, and hence
outside of the mean lightcone. \label{fig:lightcone}}
\end{figure}

Test particles can include photons, and one of the striking consequences
of gravity fluctuations can be fluctuations of the lightcone. Recall  
that the lightcone plays a crucial role in classical relativity theory.
Events which are timelike or null separated from one another can be
causally related, but those at spacelike separations cannot. Similarly,
an event horizon is a null surface which separates causally disjoint
regions of spacetime. This rigid separation cannot be maintained when
the spacetime fluctuates. A simple way to have spacetime fluctuations
is with a bath of gravitons in a nonclassical state, such as a squeezed 
vacuum state\cite{F95,FS96}. 
Here the mean spacetime geometry is almost flat, apart
from effects of the averaged stress tensor of the gravitons, but
exhibits large fluctuations around this mean. These will include
lightcone fluctuations, which will manifest themselves in varying arrival
times of pulses from a source. Consider a source and a detector, which 
are both at rest relative to the average background and separated by a
proper distance $D$, as measured in the average metric. Then the mean
flight time of pulses will be $D$, but some individual pulses will take
a longer time, and others a shorter time. A pulse which arrives in a time
less than $D$  travels outside of the lightcone of the mean spacetime,
as illustrated in Fig.~\ref{fig:lightcone}.

As noted earlier, faster than light travel can often be used to travel
backwards in time. However, there is a crucial step needed to link the two:
Lorentz invariance. One must exploit the fact that one can interchange the 
time order of spacelike separated events by changing Lorentz frames.
In the present example, Lorentz symmetry is broken by the existence of
a preferred rest frame, that of the graviton bath. Thus one cannot conclude 
that there is any problem with causality created by these lightcone 
fluctuations.            

Because an event horizon is a special case of a lightcone, there should
be horizon fluctuations in any model with spacetime geometry fluctuations.
In the case of a black hole horizon, this raises the possibility of
information leaking out of the black hole, or of the horizon fluctuations
drastically altering the semiclassical derivation of black hole evaporation.
One estimate\cite{FS97} of the magnitude of the effects of 
quantum horizon fluctuations concluded that they are too small to alter
the Hawking radiation for black holes much larger than the Planck mass.
However, other authors\cite{Sorkin,Casher} have argued for a much larger 
effect. It has also been suggested\cite{BFP00} that horizon fluctuations
might provide the new physics needed to gracefully solve the tranplanckian 
problem. This is clearly an area where more work is needed.

The passive fluctuations of gravity driven by quantum stress tensor
fluctuations are just one manifestation of stress tensor fluctuations.
They are also responsible for fluctuation forces on macroscopic bodies,
such as Casimir force fluctuations\cite{Barton,Eberlein,JR,WKF01} and 
radiation pressure fluctuations\cite{WF01}. 
This provides the possibility of an electromagnetic
analog model for passive quantum gravity. The same techniques 
are needed to define integrals of the stress tensor correlation
function in both contexts. In both cases, one needs to use a
regularization method, such as dimensional regularization\cite{FW04}
or an integration by parts. Some of the physical effect which have recently 
been studied using the latter technique are angular blurring and luminosity
fluctuations\cite{BF} of  the image of a distant source seen through a 
fluctuating spacetime.

\section{Summary}

The semiclassical theory, with quantum matter fields and a classical
gravitational field, provides a crucial link between the purely classical
theory and a more complete quantum theory of gravity. Any viable candidate
for a full quantum theory of gravity must reproduce the predictions of
semiclassical gravity in an appropriate limit. In addition, semiclassical
gravity contains a rich array of physical effects which are not found
at the classical level, including black hole evaporation, cosmological
particle creation, and negative energy density effects. The simplest 
extensions of the semiclassical theory to include spacetime fluctuations
provide another array of effects, including lightcone and horizon
fluctuations, which will have to be better understood in the context
of a more complete theory.

\section*{Acknowledgments}
\addcontentsline{toc}{section}{Acknowledgements}

I would like to thank Tom Roman for useful discussions and comments on the
manuscript.
This work was supported by the National Science Foundation under
Grant PHY-0244898.

\end{document}